\documentclass{kluwer}
\newdisplay{guess}{Conjecture}
\usepackage{graphicx}
\usepackage{url}
\graphicspath{{fig/}}

\begin{document}

% $Id: paper.tex,v 1.30 2003/06/23 09:20:05 nilshau Exp $
%|||||||||||||||||||||||||||||||||||||||||||||||||||||||||||||||||||
%             Customized Commands
%|||||||||||||||||||||||||||||||||||||||||||||||||||||||||||||||||||
%  mathematical abbreviations
%  =========================
%
% 12-feb-2003: wd introduced \BoldVec macro to get size of vectors in
%              exponents right
%
\newcommand{\BoldVec}[1]{\mathchoice%
  {\mbox{\boldmath $\displaystyle     #1$}}%
  {\mbox{\boldmath $\textstyle        #1$}}%
  {\mbox{\boldmath $\scriptstyle      #1$}}%
  {\mbox{\boldmath $\scriptscriptstyle#1$}}%
}
% math defs
\newcommand{\EQ}{\begin{equation}}
\newcommand{\EN}{\end{equation}}
\newcommand{\EQA}{\begin{eqnarray}}
\newcommand{\ENA}{\end{eqnarray}}
\newcommand{\eq}[1]{(\ref{#1})}
\newcommand{\EEq}[1]{Equation~(\ref{#1})}
\newcommand{\Eq}[1]{Eq.~(\ref{#1})}
\newcommand{\Eqs}[2]{Eqs~(\ref{#1}) and~(\ref{#2})}
\newcommand{\eqs}[2]{(\ref{#1}) and~(\ref{#2})}
\newcommand{\Eqss}[2]{Eqs~(\ref{#1})--(\ref{#2})}
\newcommand{\Sec}[1]{\S\,\ref{#1}}
\newcommand{\Secs}[2]{\S\S\,\ref{#1} and~\ref{#2}}
\newcommand{\Fig}[1]{Fig.~\ref{#1}}
\newcommand{\FFig}[1]{Figure~\ref{#1}}
\newcommand{\Tab}[1]{Table~\ref{#1}}
\newcommand{\Figs}[2]{Figures~\ref{#1} and \ref{#2}}
\newcommand{\Tabs}[2]{Tables~\ref{#1} and \ref{#2}}
\newcommand{\bra}[1]{\langle #1\rangle}
\newcommand{\bbra}[1]{\left\langle #1\right\rangle}
\newcommand{\mean}[1]{\overline #1}
\newcommand{\meanB}{\overline{B}}
\newcommand{\meanAA}{\overline{\mbox{\boldmath $A$}}}
\newcommand{\meanBB}{\overline{\mbox{\boldmath $B$}}}
\newcommand{\meanJJ}{\overline{\mbox{\boldmath $J$}}}
\newcommand{\meanuu}{\overline{\mbox{\boldmath $u$}}}
\newcommand{\meanAB}{\overline{\mbox{\boldmath $A\cdot B$}}}
\newcommand{\meanAoBo}{\overline{\mbox{\boldmath $A_0\cdot B_0$}}}
\newcommand{\meanApoBpo}{\overline{\mbox{\boldmath $A'_0\cdot B'_0$}}}
\newcommand{\meanApBp}{\overline{\mbox{\boldmath $A'\cdot B'$}}}
\newcommand{\meanuxB}{\overline{\mbox{\boldmath $\delta u\times \delta B$}}}
\newcommand{\mod}[1]{\mid\!\!#1\!\!\mid}
\newcommand{\chk}[1]{[{\em check: #1}]}
%\newcommand{\inst}[1]{$^{#1}$}
%
% tilde
%
\newcommand{\teps}{\tilde{\epsilon} {}}
%
%  unit vectors
%
\newcommand{\nnn}{\hat{\mbox{\boldmath $n$}} {}}
\newcommand{\vvv}{\hat{\mbox{\boldmath $v$}} {}}
\newcommand{\rr}{\hat{\mbox{\boldmath $r$}} {}}
\newcommand{\xxx}{\hat{\mbox{\boldmath $x$}} {}}
\newcommand{\yyy}{\hat{\mbox{\boldmath $y$}} {}}
\newcommand{\zz}{\hat{\mbox{\boldmath $z$}} {}}
\newcommand{\pp}{\hat{\mbox{\boldmath $\phi$}} {}}
\newcommand{\ttt}{\hat{\mbox{\boldmath $\theta$}} {}}
\newcommand{\OOO}{\hat{\mbox{\boldmath $\Omega$}} {}}
\newcommand{\ooo}{\hat{\mbox{\boldmath $\omega$}} {}}
\newcommand{\BBBB}{\hat{\mbox{\boldmath $B$}} {}}
%
%  vectors
%
\newcommand{\gggg}{\BoldVec{g} {}}
\newcommand{\ddd}{\BoldVec{d} {}}
\newcommand{\rrr}{\BoldVec{r} {}}
\newcommand{\xx}{\BoldVec{x}{}}
\newcommand{\yy}{\BoldVec{y} {}}
\newcommand{\zzz}{\BoldVec{z} {}}
\newcommand{\uu}{\BoldVec{u} {}}
\newcommand{\vv}{\BoldVec{v} {}}
\newcommand{\ww}{\BoldVec{w} {}}
\newcommand{\mm}{\BoldVec{m} {}}
\newcommand{\PP}{\BoldVec{P} {}}
\newcommand{\QQ}{\BoldVec{Q} {}}
\newcommand{\UU}{\BoldVec{U} {}}
\newcommand{\bb}{\BoldVec{b} {}}
\newcommand{\qq}{\BoldVec{q} {}}
\newcommand{\BB}{\BoldVec{B} {}}
\newcommand{\HH}{\BoldVec{H} {}}
\newcommand{\II}{\BoldVec{I} {}}
\newcommand{\AAA}{\BoldVec{A} {}}
\newcommand{\aaa}{\BoldVec{a} {}}
\newcommand{\aaaa}{\BoldVec{a} {}} %(convert aaa -> aaaa, compatibility problem)
\newcommand{\eee}{\BoldVec{e} {}}
\newcommand{\jj}{\BoldVec{j} {}}
\newcommand{\JJ}{\BoldVec{J} {}}
\newcommand{\nn}{\BoldVec{n} {}}
\newcommand{\ee}{\BoldVec{e} {}}
\newcommand{\ff}{\BoldVec{f} {}}
\newcommand{\EE}{\BoldVec{E} {}}
\newcommand{\FF}{\BoldVec{F} {}}
\newcommand{\TT}{\BoldVec{T} {}}
\newcommand{\CC}{\BoldVec{C} {}}
\newcommand{\KK}{\BoldVec{K} {}}
\newcommand{\MM}{\BoldVec{M} {}}
\newcommand{\GG}{\BoldVec{G} {}}
\newcommand{\kk}{\BoldVec{k} {}}
\newcommand{\SSS}{\BoldVec{S} {}}
\newcommand{\grav}{\BoldVec{g} {}}
\newcommand{\nab}{\BoldVec{\nabla} {}}
\newcommand{\OO}{\BoldVec{\Omega} {}}
\newcommand{\oo}{\BoldVec{\omega} {}}
\newcommand{\LL}{\BoldVec{\Lambda} {}}
\newcommand{\llambda}{\BoldVec{\lambda} {}}
\newcommand{\pomega}{\BoldVec{\varpi} {}}
%
%  correlation tensors
%
\newcommand{\SSSS}{\mbox{\boldmath ${\sf S}$} {}}
\newcommand{\BBB}{\mbox{\boldmath ${\cal B}$} {}}
\newcommand{\emf}{\mbox{\boldmath ${\cal E}$} {}}
\newcommand{\FFF}{\mbox{\boldmath ${\cal F}$} {}}
\newcommand{\GGG}{\mbox{\boldmath ${\cal G}$} {}}
\newcommand{\HHH}{\mbox{\boldmath ${\cal H}$} {}}
\newcommand{\QQQ}{\mbox{\boldmath ${\cal Q}$} {}}
\newcommand{\GGGG}{{\bf G} {}}
%
%  operators  (roman)
%
\newcommand{\grad}{{\rm grad} \, {}}
\newcommand{\curl}{{\rm curl} \, {}}
\newcommand{\dive}{{\rm div}  \, {}}
\newcommand{\Dive}{{\rm Div}  \, {}}
\newcommand{\DD}{{\rm D} {}}
\newcommand{\dd}{{\rm d} {}}
\newcommand{\const}{{\rm const}  {}}
\newcommand{\crit}{{\rm crit}  {}}
\def\degr{\hbox{$^\circ$}}
\def\la{\mathrel{\mathchoice {\vcenter{\offinterlineskip\halign{\hfil
$\displaystyle##$\hfil\cr<\cr\sim\cr}}}
{\vcenter{\offinterlineskip\halign{\hfil$\textstyle##$\hfil\cr<\cr\sim\cr}}}
{\vcenter{\offinterlineskip\halign{\hfil$\scriptstyle##$\hfil\cr<\cr\sim\cr}}}
{\vcenter{\offinterlineskip\halign{\hfil$\scriptscriptstyle##$\hfil\cr<\cr\sim\cr}}}}}
\def\ga{\mathrel{\mathchoice {\vcenter{\offinterlineskip\halign{\hfil
$\displaystyle##$\hfil\cr>\cr\sim\cr}}}
{\vcenter{\offinterlineskip\halign{\hfil$\textstyle##$\hfil\cr>\cr\sim\cr}}}
{\vcenter{\offinterlineskip\halign{\hfil$\scriptstyle##$\hfil\cr>\cr\sim\cr}}}
{\vcenter{\offinterlineskip\halign{\hfil$\scriptscriptstyle##$\hfil\cr>\cr\sim\cr}}}}}
%
%  numbers
%
\def\Ta{\mbox{\rm Ta}}
\def\Ra{\mbox{\rm Ra}}
\def\Ma{\mbox{\rm Ma}}
\def\Roo{\mbox{\rm Ro}^{-1}}
\def\Pra{\mbox{\rm Pr}}
\def\Pran{\mbox{\rm Pr}}
\def\Pm{\mbox{\rm Pr}_M}
\def\Rm{\mbox{\rm Re}_M}
\def\Rey{\mbox{\rm Re}}
\def\Pe{\mbox{\rm Pe}}
\newcommand{\ea}{{\em et al. }}
\newcommand{\eaa}{{\em et al. }}
\def\half{{\textstyle{1\over2}}}
\def\threehalf{{\textstyle{3\over2}}}
\def\onethird{{\textstyle{1\over3}}}
\def\twothird{{\textstyle{2\over3}}}
\def\fourthird{{\textstyle{4\over3}}}
\def\quarter{{\textstyle{1\over4}}}
\newcommand{\W}{\,{\rm W}}
\newcommand{\V}{\,{\rm V}}
\newcommand{\kV}{\,{\rm kV}}
\newcommand{\T}{\,{\rm T}}
\newcommand{\G}{\,{\rm G}}
\newcommand{\Hz}{\,{\rm Hz}}
\newcommand{\kHz}{\,{\rm kHz}}
\newcommand{\kG}{\,{\rm kG}}
\newcommand{\K}{\,{\rm K}}
\newcommand{\g}{\,{\rm g}}
\newcommand{\s}{\,{\rm s}}
\newcommand{\ms}{\,{\rm ms}}
\newcommand{\cm}{\,{\rm cm}}
\newcommand{\m}{\,{\rm m}}
\newcommand{\km}{\,{\rm km}}
\newcommand{\kms}{\,{\rm km/s}}
\newcommand{\kg}{\,{\rm kg}}
\newcommand{\Mm}{\,{\rm Mm}}
\newcommand{\pc}{\,{\rm pc}}
\newcommand{\kpc}{\,{\rm kpc}}
\newcommand{\yr}{\,{\rm yr}}
\newcommand{\Myr}{\,{\rm Myr}}
\newcommand{\Gyr}{\,{\rm Gyr}}
\newcommand{\erg}{\,{\rm erg}}
\newcommand{\mol}{\,{\rm mol}}
\newcommand{\dyn}{\,{\rm dyn}}
\newcommand{\J}{\,{\rm J}}
\newcommand{\RM}{\,{\rm RM}}
\newcommand{\EM}{\,{\rm EM}}
\newcommand{\AU}{\,{\rm AU}}
\newcommand{\A}{\,{\rm A}}
%
%  journals
%
\newcommand{\yan}[3]{, Astron. Nachr. {\bf #2}, #3 (#1).}
\newcommand{\yact}[3]{, Acta Astron. {\bf #2}, #3 (#1).}
\newcommand{\yana}[3]{, Astron. Astrophys. {\bf #2}, #3 (#1).}
\newcommand{\yanas}[3]{, Astron. Astrophys. Suppl. {\bf #2}, #3 (#1).}
\newcommand{\yanal}[3]{, Astron. Astrophys. Lett. {\bf #2}, #3 (#1).}
\newcommand{\yass}[3]{, Astrophys. Spa. Sci. {\bf #2}, #3 (#1).}
\newcommand{\ysci}[3]{, Science {\bf #2}, #3 (#1).}
\newcommand{\ysph}[3]{, Solar Phys. {\bf #2}, #3 (#1).}
\newcommand{\yjetp}[3]{, Sov. Phys. JETP {\bf #2}, #3 (#1).}
\newcommand{\yspd}[3]{, Sov. Phys. Dokl. {\bf #2}, #3 (#1).}
\newcommand{\ysov}[3]{, Sov. Astron. {\bf #2}, #3 (#1).}
\newcommand{\ysovl}[3]{, Sov. Astron. Letters {\bf #2}, #3 (#1).}
\newcommand{\ymn}[3]{, Monthly Notices Roy. Astron. Soc. {\bf #2}, #3 (#1).}
\newcommand{\yqjras}[3]{, Quart. J. Roy. Astron. Soc. {\bf #2}, #3 (#1).}
\newcommand{\ynat}[3]{, Nature {\bf #2}, #3 (#1).}
\newcommand{\sjfm}[2]{, J. Fluid Mech., submitted (#1).}
\newcommand{\pjfm}[2]{, J. Fluid Mech., in press (#1).}
\newcommand{\yjfm}[3]{, J. Fluid Mech. {\bf #2}, #3 (#1).}
\newcommand{\ypr}[3]{, Phys.\ Rev.\ {\bf #2}, #3 (#1).}
\newcommand{\yprl}[3]{, Phys.\ Rev.\ Lett.\ {\bf #2}, #3 (#1).}
\newcommand{\yepl}[3]{, Europhys. Lett. {\bf #2}, #3 (#1).}
\newcommand{\pcsf}[2]{, Chaos, Solitons \& Fractals, in press (#1).}
\newcommand{\ycsf}[3]{, Chaos, Solitons \& Fractals{\bf #2}, #3 (#1).}
\newcommand{\yprs}[3]{, Proc. Roy. Soc. Lond. {\bf #2}, #3 (#1).}
\newcommand{\yptrs}[3]{, Phil. Trans. Roy. Soc. {\bf #2}, #3 (#1).}
\newcommand{\yjcp}[3]{, J. Comp. Phys. {\bf #2}, #3 (#1).}
\newcommand{\yjgr}[3]{, J. Geophys. Res. {\bf #2}, #3 (#1).}
\newcommand{\ygrl}[3]{, Geophys. Res. Lett. {\bf #2}, #3 (#1).}
\newcommand{\yobs}[3]{, Observatory {\bf #2}, #3 (#1).}
\newcommand{\yaj}[3]{, Astronom. J. {\bf #2}, #3 (#1).}
\newcommand{\yapj}[3]{, Astrophys. J. {\bf #2}, #3 (#1).}
\newcommand{\yapjs}[3]{, Astrophys. J. Suppl. {\bf #2}, #3 (#1).}
\newcommand{\yapjl}[3]{, Astrophys. J. {\bf #2}, #3 (#1).}
\newcommand{\ypp}[3]{, Phys. Plasmas {\bf #2}, #3 (#1).}
\newcommand{\ypasj}[3]{, Publ. Astron. Soc. Japan {\bf #2}, #3 (#1).}
\newcommand{\ypac}[3]{, Publ. Astron. Soc. Pacific {\bf #2}, #3 (#1).}
\newcommand{\yannr}[3]{, Ann. Rev. Astron. Astrophys. {\bf #2}, #3 (#1).}
\newcommand{\yanar}[3]{, Astron. Astrophys. Rev. {\bf #2}, #3 (#1).}
\newcommand{\yanf}[3]{, Ann. Rev. Fluid Dyn. {\bf #2}, #3 (#1).}
\newcommand{\ypf}[3]{, Phys. Fluids {\bf #2}, #3 (#1).}
\newcommand{\yphy}[3]{, Physica {\bf #2}, #3 (#1).}
\newcommand{\ygafd}[3]{, Geophys. Astrophys. Fluid Dynam. {\bf #2}, #3 (#1).}
\newcommand{\yzfa}[3]{, Zeitschr. f. Astrophys. {\bf #2}, #3 (#1).}
\newcommand{\yptp}[3]{, Progr. Theor. Phys. {\bf #2}, #3 (#1).}
\newcommand{\yjour}[4]{, #2 {\bf #3}, #4 (#1).}
\newcommand{\pjour}[3]{, #2, in press (#1).}
\newcommand{\sjour}[3]{, #2, submitted (#1).}
\newcommand{\yprep}[2]{, #2, preprint (#1).}
\newcommand{\pproc}[3]{, (ed. #2), #3 (#1) (to appear).}
\newcommand{\yproc}[4]{, (ed. #3), pp. #2. #4 (#1).}
\newcommand{\ybook}[3]{ {\em #2}. #3 (#1).}

\begin{article}
\begin{opening}

\title{High resolution simulations of nonhelical MHD turbulence}

\author{N.E.L. \surname{Haugen}
  \email{nils.haugen@phys.ntnu.no}}
  \institute{Department of Physics, The Norwegian University of Science
  and Technology, H{\o}yskoleringen 5, N-7034 Trondheim, Norway}
\author{A. \surname{Brandenburg}
  \email{brandenb@nordita.dk}}
  \institute{NORDITA, Blegdamsvej 17, DK-2100 Copenhagen \O, Denmark}
\author{W. \surname{Dobler}
  \email{Wolfgang.Dobler@kis.uni-freiburg.de}}
  \institute{Kiepenheuer-Institut f\"ur Sonnenphysik,
  Sch\"oneckstra\ss{}e 6, D-79104 Freiburg, Germany}

\runningtitle{High resolution simulations of nonhelical MHD turbulence}

\date{\today,~ $ $Revision: 1.30 $ $}

%\begin{ao}
%Kluwer Prepress Department\\
%P.O. Box 990\\
%3300 AZ Dordrecht\\
%The Netherlands
%end{ao} 

\begin{abstract} 
According to the kinematic theory of nonhelical dynamo action the
magnetic energy spectrum increases with wavenumber and peaks at the
resistive cutoff wavenumber. It has previously been argued that even in the dynamical
case the magnetic energy peaks at the resistive scale.
Using high resolution simulations (up to $1024^3$ meshpoints)
with no large scale imposed field we show
that the magnetic energy peaks at a wavenumber that is independent
of the magnetic Reynolds number and about 5 times larger than the
forcing wavenumber. Throughout the inertial range the spectral 
magnetic energy exceeds the kinetic energy by a factor of 2 to 3.
Both spectra are approximately parallel.
The total energy
spectrum seems to be close to $k^{-3/2}$, but there is a strong 
bottleneck effect
and it is suggested that the asymptotic spectrum is instead $k^{-5/3}$.
This is supported by the value of the second order structure function exponent
that is found to be $\zeta_2=0.70$, suggesting a $k^{-1.70}$ spectrum.
The third order structure function scaling exponent is very close to unity,
in agreement with Goldreich-Sridhar theory.

Adding an imposed field tends to suppress
the small scale magnetic field. We find that
at large scales the magnetic power spectrum follows then a $k^{-1}$
slope. When the strength of the imposed field is of the same order as
the dynamo generated field, we find almost equipartition between  the 
magnetic and kinetic power spectra.

\end{abstract}
\end{opening}

\section{Introduction}

Magnetic fields may play an important role during star formation.
Stars are generally formed in strongly magnetized regions, and the
magnetic pressure that builds up in shocks and the initial collapse
is likely to determine the detailed evolution.

Early simulations of hydromagnetic turbulence have always suggested
that the magnetic field is more intermittent than the velocity field
if the field is generated by dynamo action \cite{MFP81,KYM91}.
Furthermore, linear theory \cite{Kaz68} suggests that the magnetic
spectrum should peak at the resistive scale, and it has been
argued that this may hold even in the nonlinear regime \cite{MB02}.

On the other hand, if there is an imposed large scale field there
is no doubt that most of the magnetic energy resides at large scales
e.g. Cho \& Vishniac (2000).
The obvious question is therefore, is it really true that the cases
of dynamo-generated and imposed fields are indeed drastically different?

The purpose of this paper is to compare the case of an imposed field
with that of a dynamo-generated one.
We begin by briefly reviewing the main results of our recent paper
\cite{HBD03} where we show that in the dynamo case the magnetic
energy does {\it not} peak at the resistive scale.

\section{Equations}

Our approach is the same as in Haugen et al. (2003), which is similar to
that of Brandenburg (2001), except that the flow is now forced without
helicity. We adopt an
isothermal equation of state with a constant sound speed $c_{\rm s}$,
so the pressure $p$ is related to the density $\rho$ by
$p=\rho c_{\rm s}^2$. The equation of motion is written in the form
\EQ
{\DD\uu\over\DD t}=-c_{\rm s}^2\nab\ln\rho+{\JJ\times\BB\over\rho}
+\FF_{\rm visc}+\ff,
\label{dudt}
\EN
where $\DD/\DD t=\partial/\partial t+\uu\cdot\nab$ is the advective
derivative, $\JJ=\nab\times\BB/\mu_0$ is the current density, $\mu_0$
is the vacuum permeability, $\FF_{\rm visc}$ is the viscous force,
and $\ff$ is a
random forcing function that consists of non-helical
plane waves. The continuity equation is
written in terms of the logarithmic density,
\EQ
{\DD\ln\rho\over\DD t}=-\nab\cdot\uu,
\EN
and the induction equation is solved in terms of the magnetic vector
potential $\AAA$, where $\BB=\nab\times\AAA$, so
\EQ
{\partial\AAA\over\partial t}=\uu\times\BB+\eta\nabla^2\AAA,
\label{dAdt}
\EN
where $\eta=\mbox{const}$ is the magnetic diffusivity.
We use periodic boundary conditions in all three directions
for all variables. 

The solutions are characterized by the kinetic and magnetic Reynolds
numbers,
\EQ
\mbox{Re}=u_{\rm rms}/(\nu k_{\rm f}),\quad
R_{\rm m}=u_{\rm rms}/(\eta k_{\rm f}),
\EN
respectively.
The ratio of the two is the magnetic Prandtl number,
\EQ
P_{\rm m}=\nu/\eta=R_{\rm m}/\mbox{Re}.
\EN

We use the Pencil Code\footnote{\url{
http://www.nordita.dk/data/brandenb/pencil-code}}
which is a cache and memory efficient grid based high order code (sixth
order in space and third order in time) for solving the compressible
MHD equations.

\section{Results}

\subsection{The peak of the magnetic power spectrum}

\begin{figure}[t!]\centering\includegraphics[width=0.50\textwidth]
{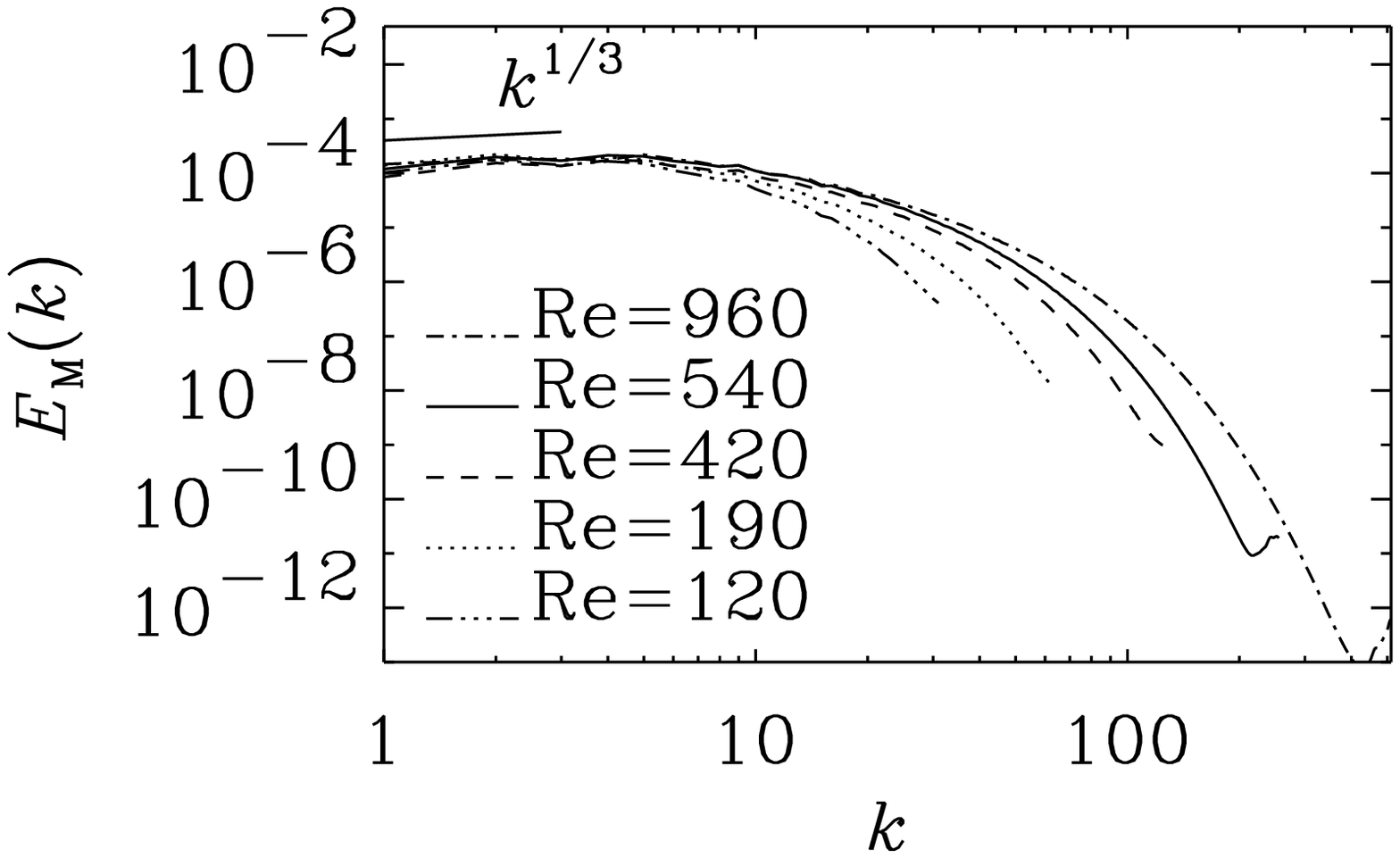}\includegraphics[width=0.50\textwidth]
{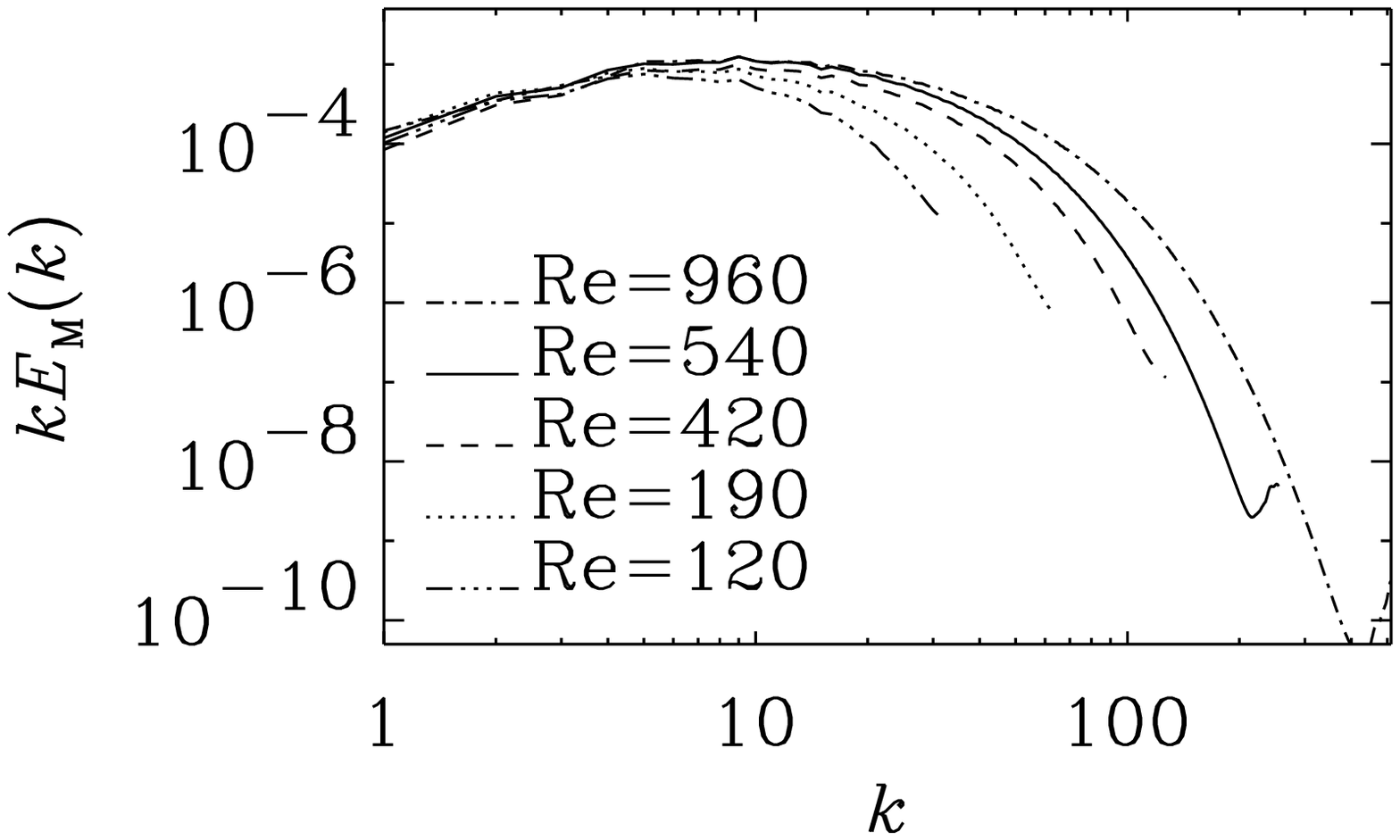}
\caption{
{\it Left}:
Magnetic power spectra for all our runs. We see that the peak of the 
magnetic power spectrum is at $k=5$ for all Re.
{\it Right}:
The peak of $B_{{\rm rms},l^2}$ is found at $k \approx 9$.
}\label{mag_comp}\end{figure}

We have run simulations with up to $1024^3$ meshpoints in order to show that
the magnetic energy spectrum does not peak at the resistive scale,
as has previously been claimed \cite{MB02}.
In the left panel of figure \ref{mag_comp} 
we see that the peak of the spectrum is around
$k=5$ for all our runs, i.e. it is independent of Re. We also
note that we find a $k^{1/3}$
slope for small values of $k$ \cite{Bat50}.

A more stringent measure is to look at the magnetic energy per unit
logarithmic wavenumber interval, $kE_M(k)$, which would be flat
if the contributions from small and large wavenumbers was equal.
This is shown in the right hand panel 
of figure \ref{mag_comp}. 
We see that the peak of $kE_M(k)$ is shifted toward smaller 
scales compared to $E_M(k)$, but it is still not at the resistive scale.
We do indeed see that for the largest runs it seems to settle at 
$k \approx 10$, which is well within the inertial range.

\subsection{The inertial range} 

\begin{figure}[t!]\centering\includegraphics[width=0.80\textwidth]
{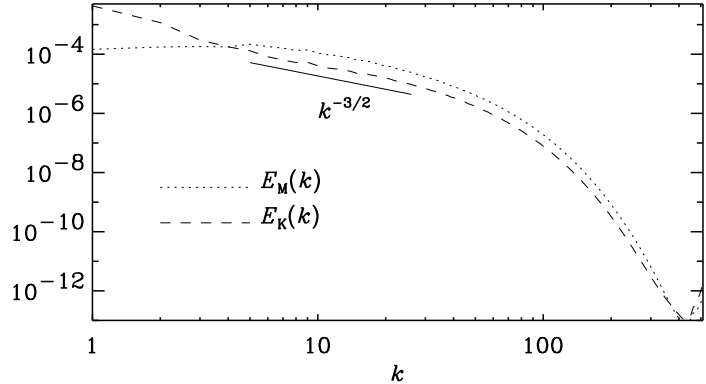}\caption{
Kinetic and magnetic power spectrum of our largest run with $1024^3$
mesh points.
}\label{power1024a}\end{figure}

\begin{figure}[t!]\centering
\includegraphics[width=0.50\textwidth]
{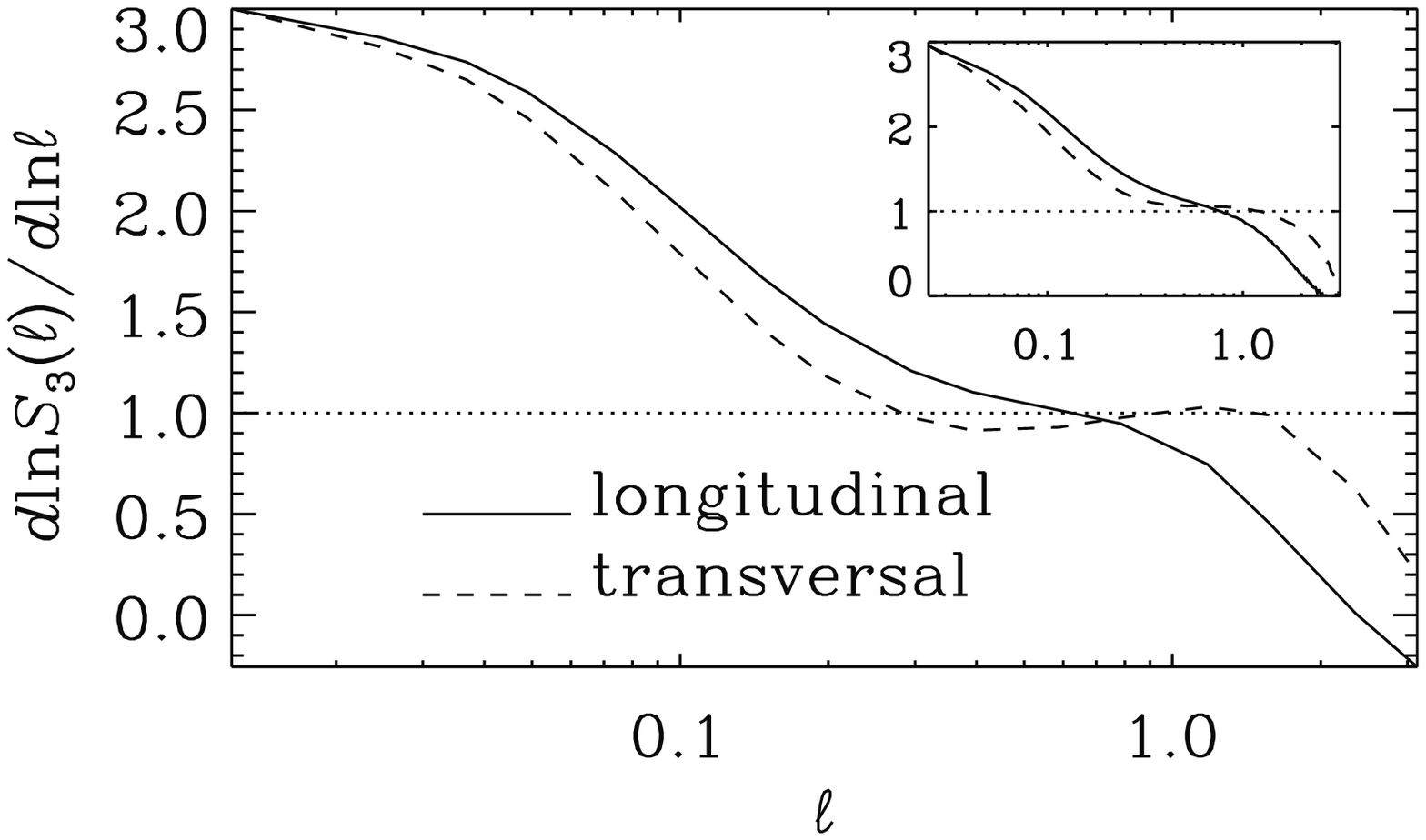}\includegraphics[width=0.50\textwidth]
{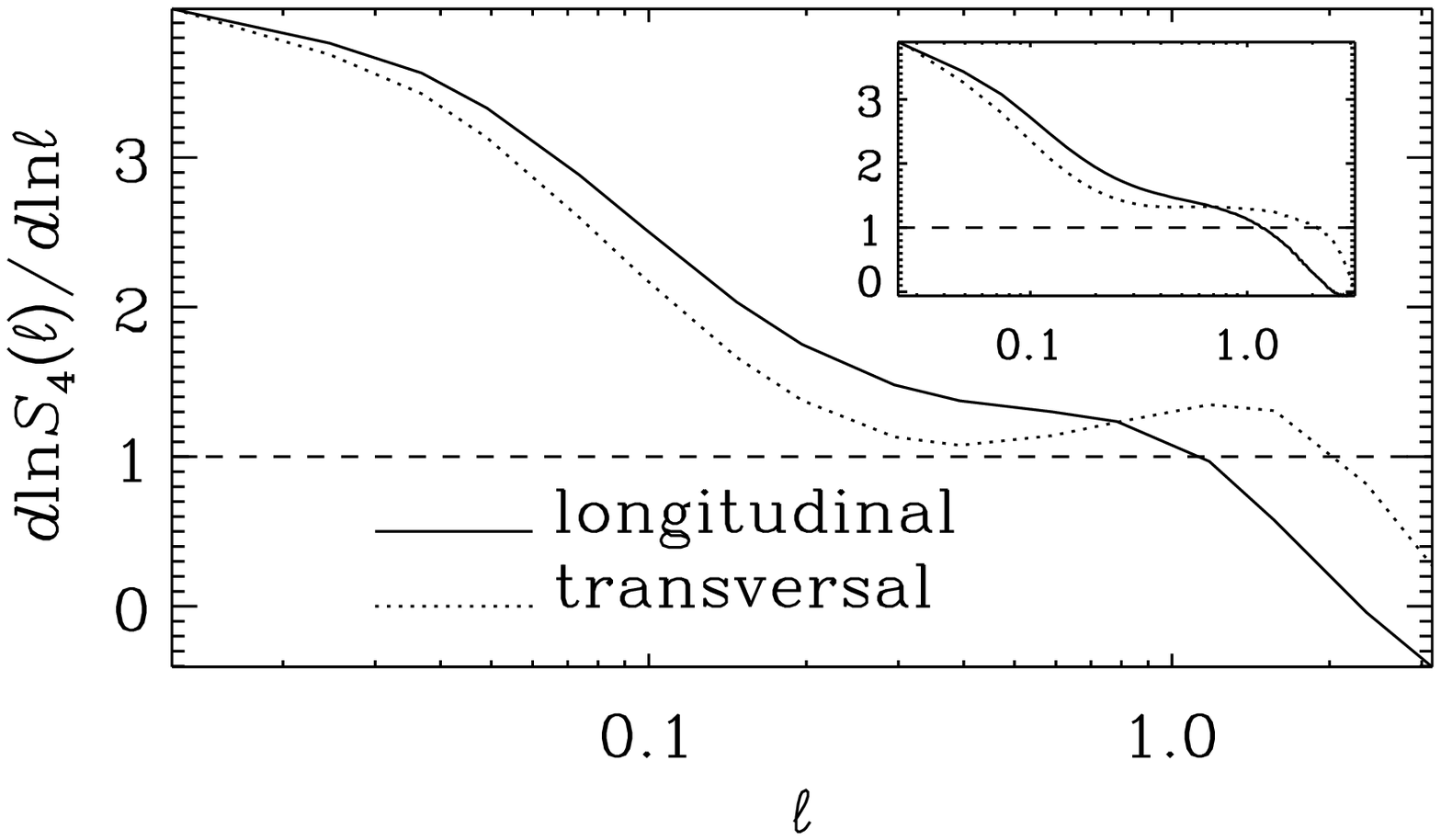}
\caption{
{\it Left:}
logarithmic derivative of third order structure function. The inset is
for a run with $256^3$ mesh points, while the large plot is for a run with
$512^3$ mesh points. The result is consistent with $\zeta_3=1$
{\it Right:}
logarithmic derivative of fourth order structure function.
We see that this is clearly {\it not} compatible with $\zeta_4=1$
}\label{moment512_256}\end{figure}

\begin{figure}[t!]\centering\includegraphics[width=0.90\textwidth]
{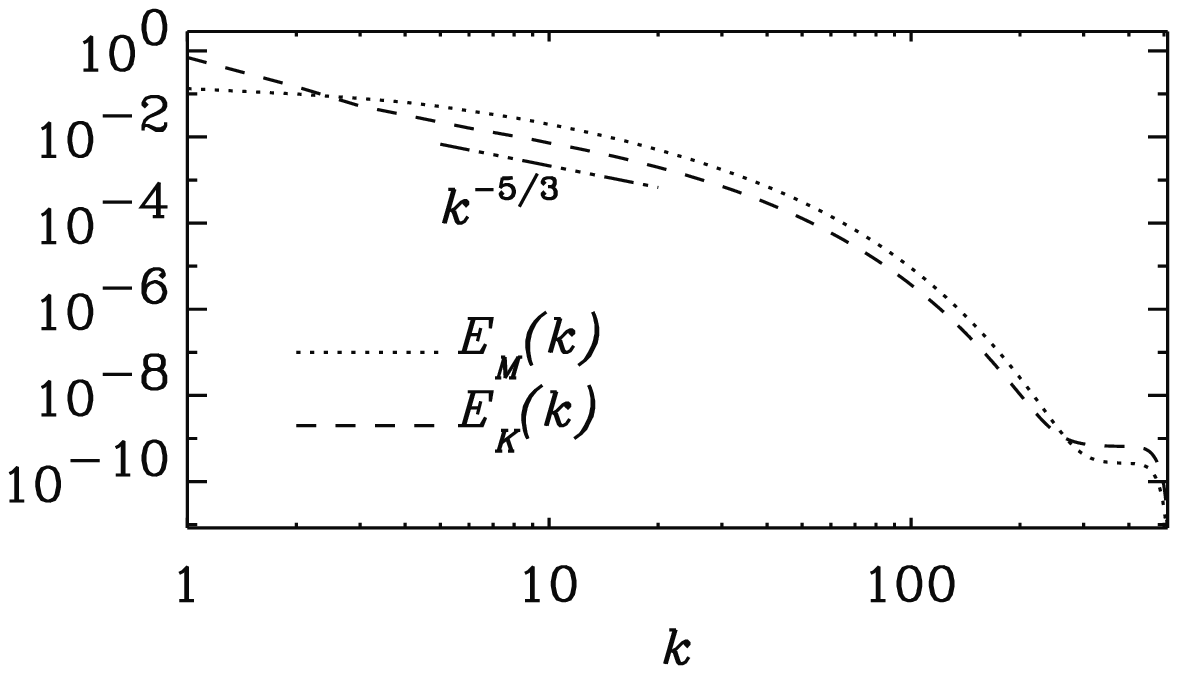}\caption{
one-dimensional power spectra of our largest run. We see here that the
inertial range is consistent with a $k^{-5/3}$ slope.
}\label{power1024a_1d}\end{figure}

In figure \ref{power1024a} we see that there seem to be a clear inertial
range for $7 \lesssim k \lesssim 25$ where $E_M(k)$ and $E_K(k)$ 
are parallel and have a
slope of $k^{-3/2}$. The $k^{-3/2}$ slope is suggestive of
the Iroshnikov (1963) \& Kraichnan (1965) (IK) 
theory, and may seem incompatible with
the Goldreich \& Sridhar (1995) (GS) theory. 
We also note that in the inertial range the
fraction of the magnetic and kinetic energy seem to be saturated at;
$E_M(k)/E_K(k)\approx 2.3$.

Knowing that IK theory 
predicts that the fourth order structure function scales linearly, while
GS theory predicts linear scaling for the third order structure function,
we now calculate the logarithmic derivatives for these structure functions; 
see figure \ref{moment512_256}. From these
plots we see that the IK theory cannot be correct since the 
fourth order structure function is clearly steeper than linear. The third
order structure function on the other hand scales linearly.

The second order structure function scaling exponent 
of the Elsasser variable, $\zzz^\pm=\uu\pm\BB/\sqrt{\rho \mu}$, 
is also indicative of
GS theory being applicable since we find  $\zeta_2=0.7$,
which imply $E_T(k)=E_M(k)+E_K(k) \propto k^{-(1+\zeta_2)}=k^{-1.7}$. 

As we have argued earlier \cite{HBD03},
the reason that figure \ref{power1024a} shows a $k^{-3/2}$ 
inertial range is that there is a strong bottleneck effect
\cite{Fal94}. It turns out that this 
bottleneck is much stronger in three-dimensional power spectra than
in one-dimensional ones \cite{DHYB03}. We therefore plot in figure
\ref{power1024a_1d} the
one-dimen\-sio\-nal counterpart of figure \ref{power1024a} in figure
\ref{power1024a_1d}. Here we see that the inertial range does indeed
has a slope close to $-5/3$, as suggested by our previous findings 
from the structure functions.
From this we conclude that also the three-dimensional power spectra will
show a $k^{-5/3}$ inertial range away from the diffusive subrange.

\subsection{Imposed magnetic field}

\begin{figure}[t!]\centering\includegraphics[width=0.50\textwidth]
{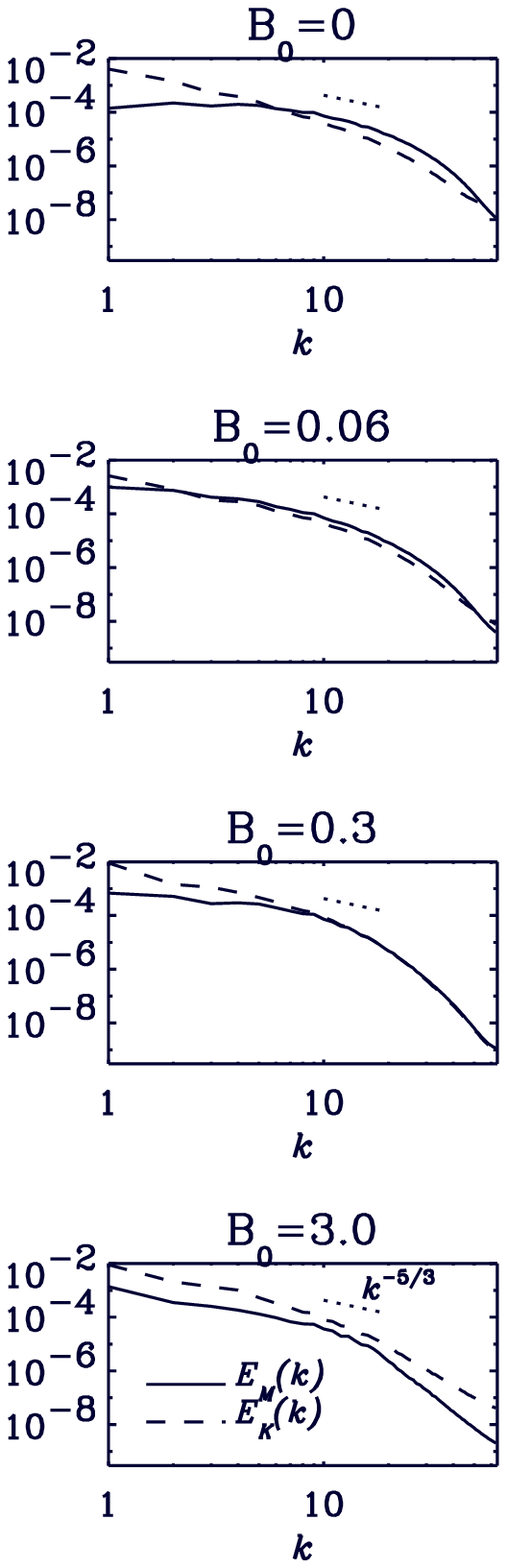}\includegraphics[width=0.50\textwidth]
{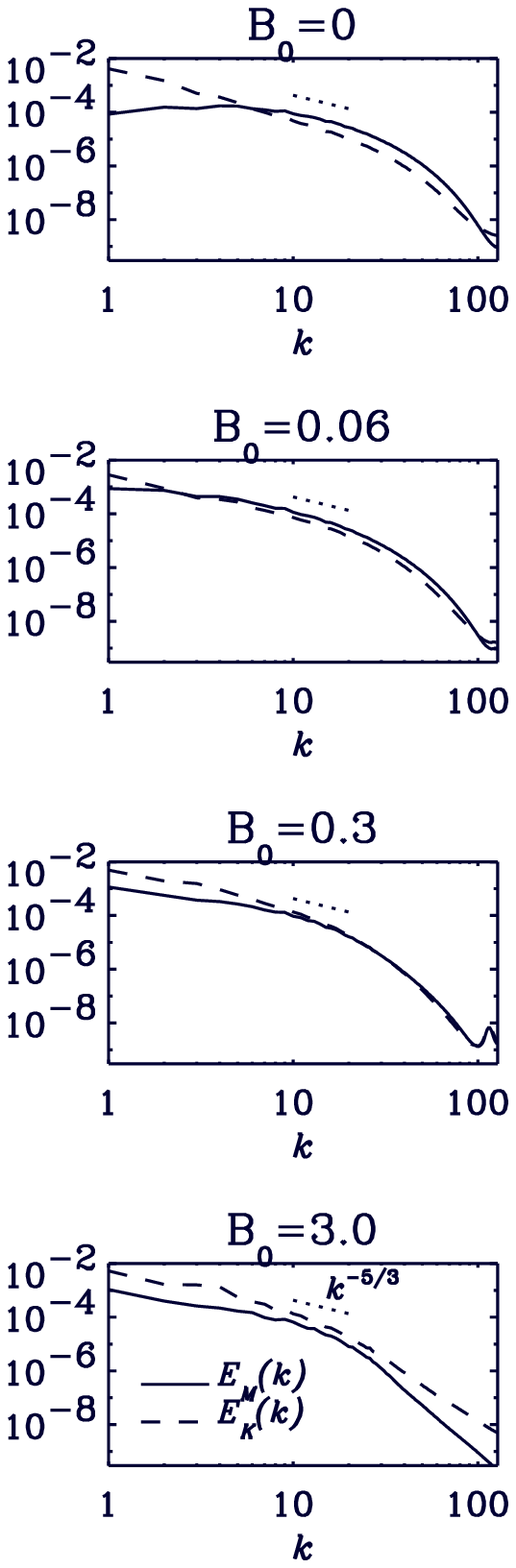}
\caption{{\it Left:}
power spectra for simulations with imposed fields of different strengths.
$128^3$ meshpoints.
{\it Right:}
power spectra for simulations with imposed fields of different strengths.
$256^3$ meshpoints.
}\label{power_comp_externalB}\end{figure}

\begin{figure}[t!]\centering\includegraphics[width=0.90\textwidth]
{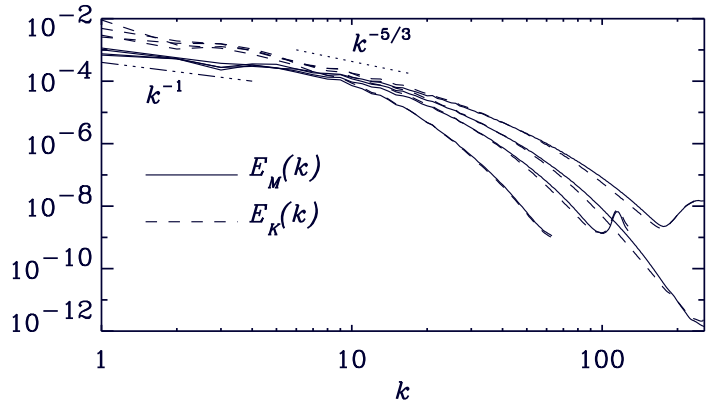}\caption{
Power spectra for runs with different Re, and with $B_0=0.3$.
We see a $k^{-1}$ for for $1<k<4$, and we also start to see the
appearance of an $k^{-5/3}$ inertial range beginning at $k \approx 8$.
}\label{power_comp_externalB_03_in_one}\end{figure}

\begin{figure}[t!]\centering\includegraphics[width=1.0\textwidth]
{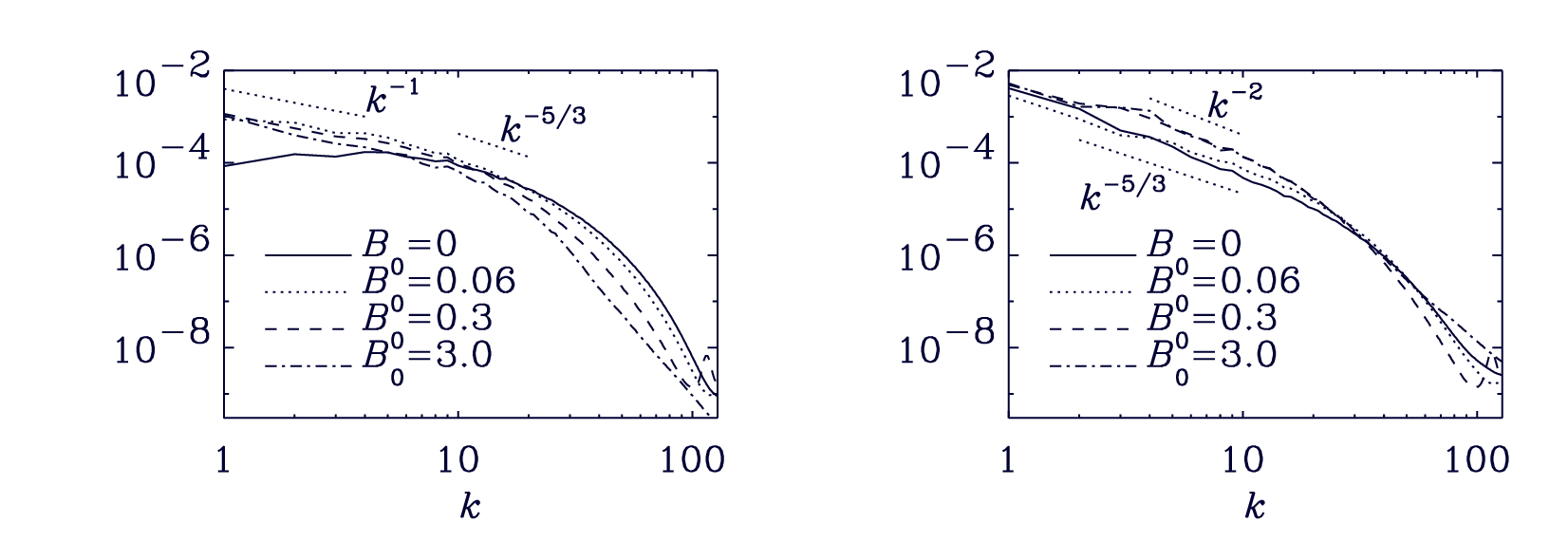}
\caption{Power spectra for runs with different imposed fields.
{\it Left:}
magnetic power spectra; we see that the the larger the external field
the more the magnetic field at small scales is suppressed.
{\it Right:}
kinetic power spectra
}\label{power_comp_externalB_256_in_one}\end{figure}

Until now we have only been looking at the case with no externally imposed
field. We therefore want to see what effect such imposed fields of varying
strengths have on the dynamo. In figure \ref{power_comp_externalB} we
plot power spectra for simulations
with $128^3$ meshpoints (left panel) and $256^3$ meshpoints (right panel) 
for different imposed fields. 
From the figure we see that having an imposed field seems to increase the
large scale magnetic field at the same time as it {\it decreases} the small 
scale field. The stronger the external field the more suppressed are the 
small scale fields. We also note that with an imposed field the peak of the 
magnetic energy is found at the largest scale $(k=1)$, not at $k=5$ as in the 
case without an imposed field.
Looking at figure \ref{power_comp_externalB_03_in_one} we see that for
$1<k<4$ there seems to be a $k^{-1}$ slope for $E_M(k)$. 
Such a slope for the large scale magnetic field has
been suggested  previously \cite{RS82,KR94,BJNRST96,MG86}.
From $k \approx 6$ there seems to be a short inertial range with 
a $k^{-5/3}$ slope, as expected from GS theory.

When the strength of the imposed field is comparable to the dynamo 
generated field 
($B_0=0.06$ and $B_0=0.3$ in figure \ref{power_comp_externalB}) we 
see that there is almost equipartition between magnetic and kinetic
energy spectra, at least for the smaller scales. 
On the other hand, Cho \& Vishniac (2000) find almost perfect 
equipartition for all scales. The difference is small, and could
perhaps be explained by a difference in the forcing function.
We did check, however, that changing from a delta-correlated
forcing function to one with a renewal time comparable to the
turnover time does not resolve this relatively minor discrepancy.

\section{Discussion}
We have shown that for non helical MHD turbulence without
imposed magnetic field,
$E_M(k)$ and $kE_M(k)$ peak at $k=5$ and $k=10$, respectively, and not
at the resistive scale. 
We also find that in the inertial range $E_M(k)$ and $E_K(k)$ are 
parallel to each other, but with $E_M(k)/E_K(k)\approx 2.5$ and a slope of $k^{-5/3}$.

If we impose an external large scale field we find that $E_M(k)$ peaks at 
the box scale, and shows a $k^{-1}$ subinertial range.  In the inertial range
we find the expected $k^{-5/3}$ slope and almost equipartition between
the kinetic and magnetic power spectra, i.e. $E_M(k)/E_K(k)\approx 1$, when
the imposed field has a strength in the order of the dynamo generated field. 
An imposed large scale magnetic field therefore has the effect of increasing
the magnetic energy at large scales, but decreasing it at small scales.

\acknowledgements
Use of the supercomputers in Trondheim (Gridur), Odense (Horseshoe),
Leicester (Ukaff) and Bergen (fire) is acknowledged.

%\theendnotes

\end{article}

\begin{thebibliography}{}

\bibitem[\protect\citeauthoryear{Brandenburg}{2001}]{B01}
A. Brandenburg\yapj{2001}{550}{824}

\bibitem[\protect\citeauthoryear{Haugen et al.}{2003}]{HBD03}
N. E. L. Haugen, A. Brandenburg and W. Dobler 
Phys. Rev. Lett. , submitted, 
{\sf astro-ph/0303372}, Paper~I.

\bibitem[\protect\citeauthoryear{Goldreich \& Sridhar}{1995}]{GS95}
P. Goldreich and S. Sridhar\yapj{1995}{438}{763}

\bibitem[\protect\citeauthoryear{Maron \& Cowley}{2001}]{MC01}
J. Maron and S. C. Cowley, {\sf astro-ph/0111008} (2001).

\bibitem[\protect\citeauthoryear{Kazantsev}{1968}]{Kaz68}
A. P. Kazantsev\yjetp{1968}{26}{1031}

\bibitem[\protect\citeauthoryear{Falkovich}{1994}]{Fal94}
G. Falkovich\ypf{1994}{6}{1411}

\bibitem[\protect\citeauthoryear{Dobler et al.}{2003}]{DHYB03}
W. Dobler, N. E. L. Haugen, T. Yousef, and
A. Brandenburg, Phys. Rev. {\bf E}, submitted,
{\sf astro-ph/0303324} (2003).

\bibitem[\protect\citeauthoryear{Brandenburg etal.}{2003}]{BHD03}
A. Brandenburg, N. E. L. Haugen and W. Dobler,
{\sf astro-ph/0303371} (2003).
% NORDITA-2003-13 AP
% {Turbulence, Waves, and Instabilities in the Solar Plasma}
% {K. Petrovay}
% {Kluwer}

\bibitem[\protect\citeauthoryear{Iroshnikov}{1963}]{Iro63}
R. S. Iroshnikov\ysov{1963}{7}{566}

\bibitem[\protect\citeauthoryear{Kraichnan}{1965}]{Kra65}
R. H. Kraichnan\ypf{1965}{8}{1385}

\bibitem[\protect\citeauthoryear{She \& Leveque}{1994}]{SL94}
Z.-S. She and E. Leveque\yprl{1994}{72}{336}

\bibitem{CLV02}
J. Cho, A. Lazarian, and E. Vishniac\yapj{2002}{564}{291}

\bibitem[\protect\citeauthoryear{Batchelor}{1950}]{Bat50}
G. K. Batchelor\yprs{1950}{A201}{405}

\bibitem[\protect\citeauthoryear{Maron \& Blackman}{2002}]{MB02}
J. Maron and E. G. Blackman\yapjl{2002}{566}{L41}

\bibitem[\protect\citeauthoryear{Ruzmaikin \& Shukurov}{1982}]{RS82}
A. A. Ruzmaikin and A. M. Shukurov\yass{1982}{82}{397}

\bibitem[\protect\citeauthoryear{Cho \& Vishniac}{2000}]{CV00}
J. Cho and E. Vishniac\yapj{2000}{539}{273}

\bibitem[\protect\citeauthoryear{Meneguzzi et al.}{1981}]{MFP81}
M. Meneguzzi, U. Frisch, and A. Pouquet\yprl{1981}{47}{1060}

\bibitem[\protect\citeauthoryear{Kida et al.}{1991}]{KYM91}
S. Kida, S. Yanase, J. Mizushima\ypf{1991}{A 3}{457}

\bibitem[\protect\citeauthoryear{Brandenburg et al.}{1996}]{BJNRST96}
Brandenburg, A., Jennings, R. L., Nordlund, \AA.,
Rieutord, M., Stein, R. F., \& Tuominen, I.\yjfm{1996}{306}{325}{352}

\bibitem[\protect\citeauthoryear{Matthaeus \& Goldstein}{1986}]{MG86}
W. H. Matthaeus and M. L. Goldstein\yprl{1986}{57}{495}

\bibitem[\protect\citeauthoryear{Kleeorin \& Rogachevskii}{1994}]{KR94}
N. Kleeorin and I. Rogachevskii\ypr{1994}{E 50}{2716}

\end{thebibliography}
\end{document}